\documentclass[12pt,epsf,amssymb]{article}
\usepackage{fullpage}
\usepackage{tabularx}
\usepackage{array}
\usepackage{graphics}
\usepackage{graphicx}
\usepackage{epsfig}
\usepackage{amsmath}
\usepackage{amssymb}
\usepackage{citesort}

\def\str{\operatorname{str}}

\begin{document}

\thispagestyle{empty} 

\begin{flushright}
IJS-TP-11/04

RBRC-424

BNL-HET-04/8

KANAZAWA-04-11

MIT-CTP-3517
\end{flushright}

\begin{center}

\vspace{1cm}

{\large \bf Scalar meson in dynamical and partially quenched  }

\vspace{0.3cm}

{\large \bf two-flavor QCD: lattice results and chiral loops }

\vspace{1cm}

\centerline{S. Prelovsek$^{a,b}$, C. Dawson$^c$, T. Izubuchi$^{c,d}$, K. Orginos$^e$  and A. Soni$^f$}

\vspace{1cm}

\centerline{\small \it $^a$ Department of Physics, University of Ljubljana, Jadranska 19,  1000 Ljubljana, Slovenia}
\centerline{\small \it $^b$ Institute Jozef Stefan, Jamova 39,  1000 Ljubljana, Slovenia}
\centerline{\small \it $^c$ RIKEN-BNL Research Center, Brookhaven National Laboratory, Upton, NY 11973, USA}
\centerline{\small \it $^d$ Institute of Theoretical Physics, Kanazawa University, Ishikawa 920-1192, Japan}
\centerline{\small \it $^e$ Center for Theoretical Physics, Laboratory for Nuclear Science and Department of Physics, }
\centerline{\small \it  MIT, Cambridge, MA 02139-4307,  USA}

\centerline{\small \it $^f$  Physics Department, Brookhaven National Laboratory, Upton, NY 11973, USA}

 \vspace{2cm}

\centerline{\bf ABSTRACT}

\vspace{0.15cm}

\end{center}

This is an exploratory study of 
the  lightest non-singlet scalar $q\bar q$ state on the lattice with 
two dynamical quarks.
Domain wall fermions are used for both sea and valence quarks on
     a $16^3 \times 32$ lattice
with an inverse lattice spacing of $1.7~$GeV.  We extract the 
 scalar meson mass $1.58\pm 0.34~$GeV  
from the exponential time-dependence of the dynamical correlators with 
$m_{val}\!=\!m_{sea}$ and $N_f=2$. 
Since this statistical error-bar from dynamical correlators is rather large, 
 we analyze also the partially quenched
lattice correlators with $m_{val}\!\not =\!m_{sea}$. They are positive for 
$m_{val}\!\geq\! m_{sea}$ and  negative for $m_{val}\!<\!m_{sea}$.
In order to understand this striking effect of partial quenching, we 
derive the scalar correlator within the Partially Quenched ChPT
and find it describes lattice correlators well. 
The leading unphysical contribution in Partially Quenched ChPT 
comes from the exchange of the two pseudoscalar fields and is   
 also positive for $m_{val}\!\geq\!m_{sea}$ and negative 
for $m_{val}\!<\!m_{sea}$ at large $t$.
  After the subtraction of this  unphysical contribution from the 
partially quenched lattice correlators, the correlators are positive and 
exponentially falling.  
The resulting scalar meson mass $1.51\pm 0.19~$GeV 
from the partially quenched correlators is 
consistent with the dynamical result and has appreciably smaller error-bar.


\newpage

\section{Introduction}

The interest in the light scalar mesons has been renewed recently 
\cite{tornqvist}. 
The existence of the scalar mesons above $1$ GeV is well established 
experimentally and there are enough scalar states between $1$ GeV and $2$ GeV 
to represent the scalar $q\bar q$ nonet \cite{pdg}. The excess of one 
observed state in this region has been suggested as an 
indication for the glueball \cite{weingarten}.
The lightest iso-triplet state above $1$ GeV is $a_0(1450)$. 
The only scalar states below $1$ GeV, 
which are experimentally well established, are  
iso-triplet $a_0(980)$ and iso-singlet $f_0(980)$ \cite{pdg}. The existence 
of the complete scalar $q\bar q$ 
nonet roughly below $1$ GeV would require  another iso-singlet 
and two strange iso-doublets. The experimental evidence for the existence of 
a broad iso-singlet $\sigma$ meson around $600$ MeV is growing \cite{pdg,e791},
while the existence of the strange iso-doublet $\kappa$ 
reported in \cite{kappa} is even more controversial at present. 
 This raises a question whether the lightest scalar 
$q\bar q$ states lie below $1$ GeV or above $1$ GeV. In the latter case, 
the observed scalar states below $1$ GeV have to be interpreted as exotic 
states like $qq\bar q\bar q$ \cite{jaffe}, $\pi\pi$ or $K\bar K$ molecules, etc. 
     
In this paper we address the determination of the mass of the 
lightest scalar $q\bar q$ state with non-singlet flavor structure 
(referred to as the $a_0$ meson \cite{pdg}), the long-term goal being 
to relate this state to the  observed resonance $a_0(1450)$ or $a_0(980)$. 
We determine the mass of the $a_0$  meson using a  
lattice simulation of 
dynamical QCD ($m_{val}=m_{sea}$)  and partially 
quenched QCD  ($m_{val}\not =m_{sea}$) with $N_f\!=\!2$ degenerate sea 
quarks in both cases. Since our aim is 
$q\bar q$ state composed of the light $u$ and $d$ quarks, 
we employ Domain Wall Fermion (DWF) formalism, which has 
good chiral properties \cite{dwf}. We comment also 
on the mass of the $s\bar u$ and $s\bar d$ scalar mesons. 

While DWF formulation ought to be helpful in the long run, at present our 
numerical work has serious limitations. We are working with two dynamical 
flavors, which is not full QCD. Furthermore we have results only at one 
lattice spacing on a lattice box that is not very large and also quarks 
are relatively heavy. 
For these reasons this is an exploratory work. 
These issues can of course be improved with more computing resources. 

\vspace{0.3cm}

Before we introduce our work, we briefly review the recent lattice simulations
  of the light non-singlet scalar states. 
We quote only the statistical error-bars  on masses since 
the continuum and infinite-volume extrapolations have not been performed 
in these simulations:   

\vspace{-0.3cm}

\begin{itemize}
\item {\it Fully quenched simulations of $q\bar q$:}

\vspace{-0.2cm} 

The quenched $q\bar q$ correlator in the 
chiral limit was simulated  by Bardeen {\it et al.} 
\cite{bardeen,bardeen2} with Wilson fermions. The correlators were found 
to be negative at small quark masses, which was attributed to the similar 
mechanism as observed in the present partially quenched study. 
The effects of quenching were modelled using the Quenched Chiral 
Perturbation Theory and subtracted in order to extract the scalar 
meson mass $m_{a0}=1.326(86)$ GeV \cite{bardeen2}.

The RBC Collaboration 
simulated non-singlet and singlet scalar $q\bar q$ states
  with Domain Wall Fermions  \cite{sasa}. 
The quenching effect, which leads to negative correlators at 
small quark masses, was subtracted as in 
\cite{bardeen,bardeen2}. The result is $m_{a0}=1.43(10)$ GeV 
if only the leading  chiral loop (one bubble) is taken into account,
and $m_{a0}=1.04(7)$ GeV if next-to leading chiral corrections are included 
by resummation\footnote{The higher order chiral corrections 
${\cal O}(M_\pi^2/(4\pi f)^2)$ have smaller effect on the scalar mass in Ref. 
\cite{bardeen,bardeen2} since their value of the pseudoscalar decay 
constant $f$ is larger than the physical value.}.

Mixing of the glueball and $q\bar q$ was studied in \cite{weingarten}. The quark mass was around $m_s$ and no attempt was made to go to the chiral limit. 

\item {\it Dynamical simulations of $q\bar q$:}

\vspace{-0.2cm} 
 
The SCALAR Collaboration 
made an extensive simulation of the singlet $\bar qq$ 
state \cite{scalar_coll} and extracted also the mass of the 
non-singlet state to be $m_{a0}\sim 1.8$ GeV at $m_\pi/m_\rho\sim 0.7$. 
They consider 
this estimate as an upper bound on the mass since they 
fitted the correlators at relatively low times, where contribution 
of the excited state might be sizable.

UKQCD extracted $m_{a0}\sim 1.0(2)$ GeV from the dynamical  and the  
partially quenched  simulation of $q\bar q$ \cite{UKQCD}. Since they 
simulated   only $m_{val}\geq m_{sea}$, they did 
not observe the striking effect of partial quenching discussed below.    
For this reason  they were able to extract the scalar mass from the 
exponential time-dependence.

MILC \cite{MILC} simulated $q\bar q$ state with three dynamical flavors 
and saw an indication for the intermediate state $\pi\eta$, 
since this state is lighter than $a_0$ state at the lightest quark masses.

\item Alford and Jaffe reported an indication for the bound singlet and 
octet {\it $qq\bar q\bar q$ states} below $1$ GeV \cite{jaffe}. 
  \end{itemize}

\vspace{-0.3cm}

All simulations above employed Wilson fermions, except for RBC and MILC 
simulations, which employed Domain Wall and staggered fermions, 
respectively. 

\vspace{0.3cm}

\begin{figure}[htb!]
\begin{center}
\epsfig{file=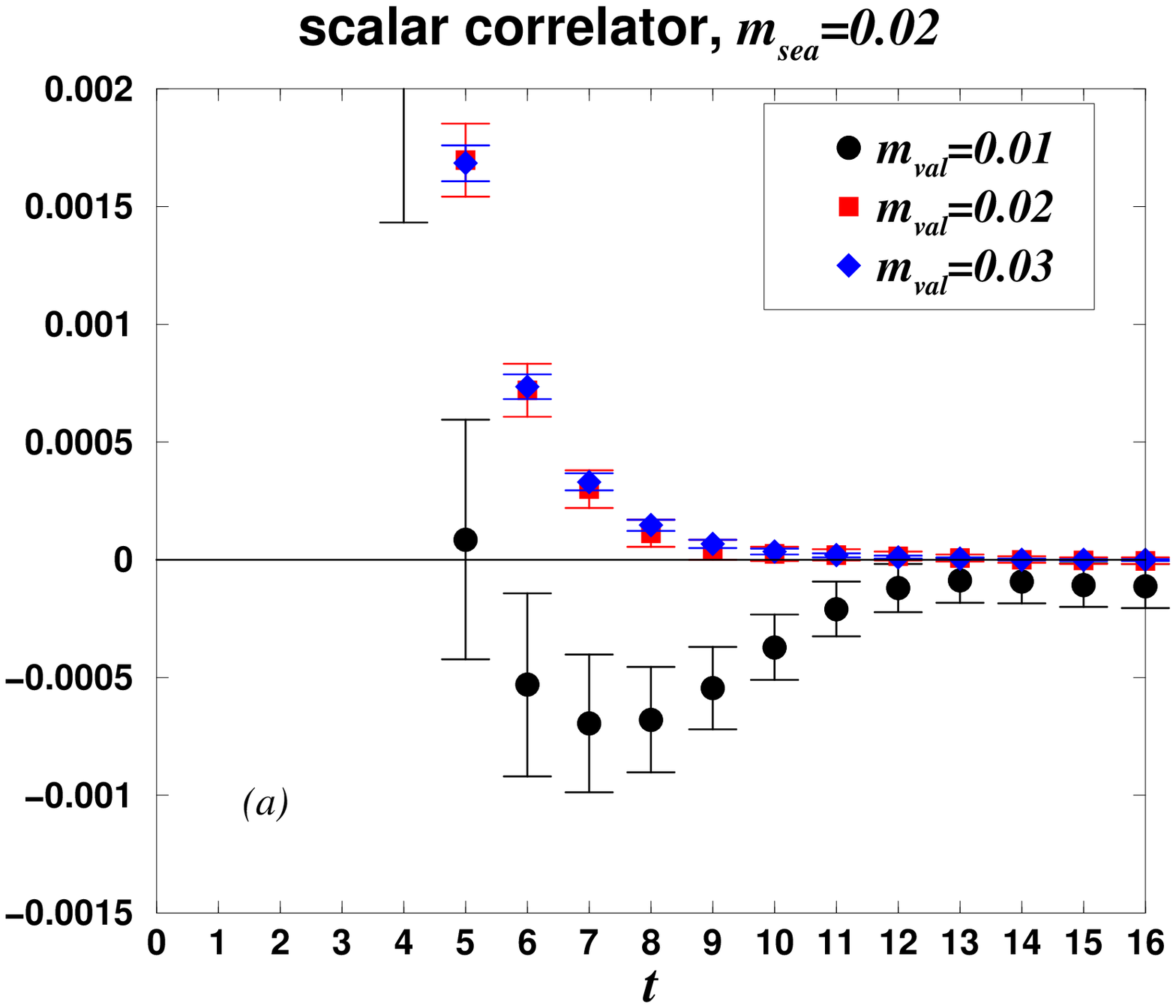,height=7cm}
$\quad$
\epsfig{file=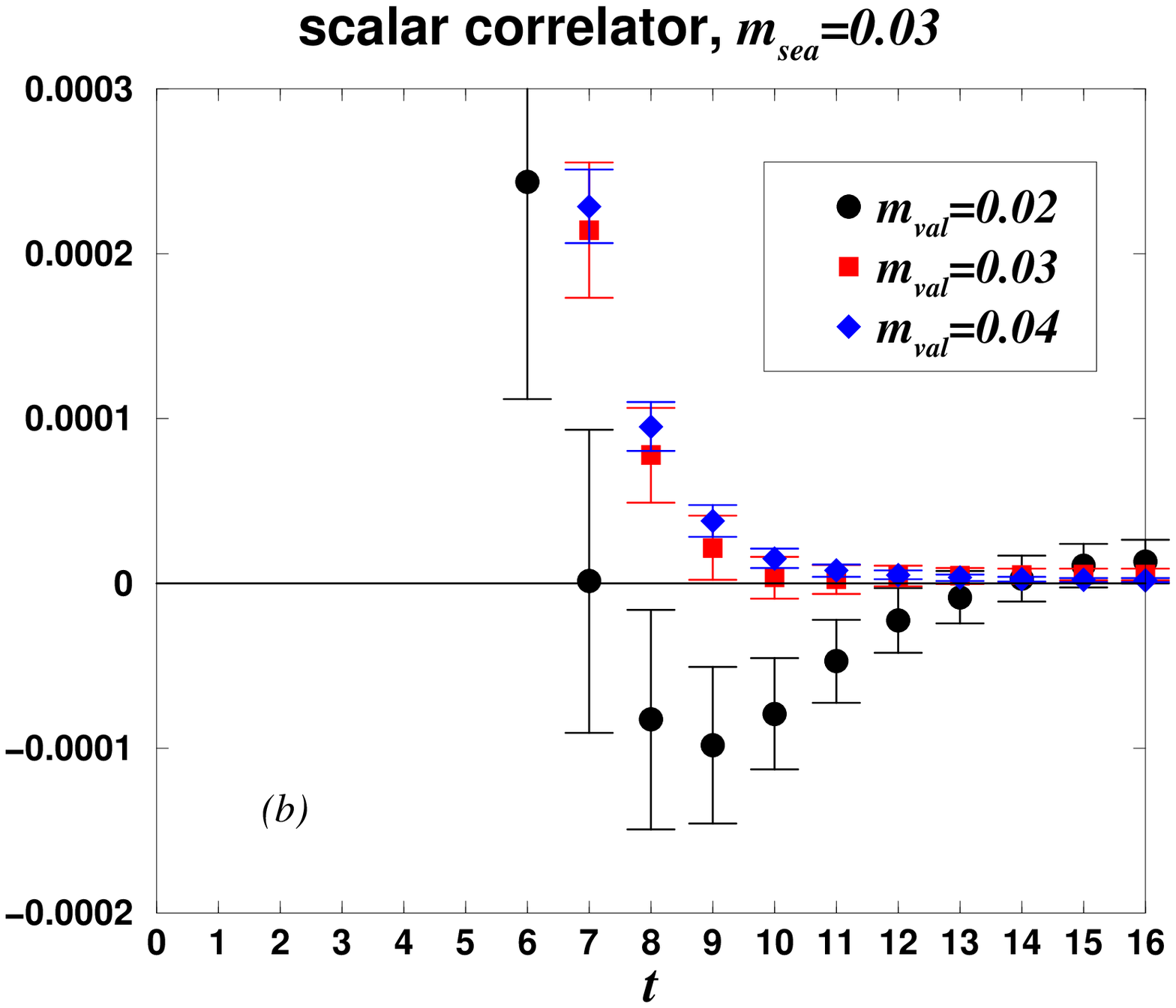,height=7cm}
\end{center}

\vspace{-0.8cm}

\caption{ \small The scalar correlators from lattice at (a) $m_{sea}=0.02$ and (b) $m_{sea}=0.03$ for various $m_{val}$.}\label{fig.1}
\end{figure}

The only simulation which employed chiral fermions to study light scalar mesons is the quenched simulation of RBC \cite{sasa}. Chiral symmetry is expected to be particularly important for the singlet scalar meson $\sigma$, 
which is intimately connected with the chiral symmetry breaking. Good 
understanding of the non-singlet correlator in the 
chiral limit is the first step toward the controlled study of the $\sigma$ meson. As already mentioned, 
the present paper presents the  dynamical simulation ($m_{sea}=m_{val}$)   
of the non-singlet $q\bar q$ correlator with Domain Wall fermions. 
We also simulate partially quenched QCD with $m_{val}\not = m_{sea}$.
Two degenerate sea quarks have the range of masses corresponding to  
$M_\pi\!\sim\! 500\! -\! 700~$MeV \cite{dyn_dwf}. The scalar correlators for 
$m_{sea}=0.02$ and $m_{sea}=0.03$ at various $m_{val}$ are shown in 
Fig. \ref{fig.1}. 
The correlators for $m_{val}\geq m_{sea}$ are positive and have more or less exponential time-dependence. On the other hand, the correlators for $m_{val}<m_{sea}$ are negative due to a striking effect of partial quenching. 
We note that the point-point correlator should be positive definite in the dynamical QCD based on unitarity, which is broken in the partially 
quenched QCD. We derive the 
effect of partial quenching on the scalar 
correlator using the Partially Quenched Chiral Perturbation Theory (PQChPT) 
\cite{pqchpt} in a finite box. 
The leading unphysical effect is due to the exchange of the 
two pseudoscalar fields, it is represented by the 
bubble diagram in Fig. \ref{fig.bubble}b and has no unknown parameters. 
We show that the bubble diagram gives a positive contribution for 
$m_{val}\geq m_{sea}$ and a negative contribution for  $m_{val}< m_{sea}$ 
at large time separations. We find that the negative lattice correlators with $m_{val}<m_{sea}$ are well described by the bubble contribution.
  This enables us to extract $m_{a0}$ in the partially quenched simulation.

\vspace{0.2cm}

The remainder of this paper is organized as follows. The  details 
about the lattice simulation are presented in section 2.
The dynamical correlators are analyzed in section 3. The resulting error 
on the scalar mass is rather large, which motivates us to 
analyze also the partially quenched correlators. 
The  partially quenched artifacts on the scalar correlator are 
derived within PQChPT in section 4 and used to analyze the partially quenched correlators in section 5.  Section 6 summarizes the conclusions on the mass of $a_0$ meson and briefly comments on mass of the $\kappa$ resonance, while section 7 summarizes the general conclusions.  

\section{Numerical simulation}

The RBC Collaboration has undertaken a large-scale simulation with 
$N_{f}\!=\!2$ flavors of dynamical Domain Wall quarks with 
degenerate masses \cite{dyn_dwf}.
This is an improvement over the quenched simulations 
and represents  an important step  toward the 
simulation of QCD with three dynamical quarks of physical masses. 
The scalar correlators were calculated on the dynamical configurations with  
the volume $N_L^3N_T=16^3 32$ and a single lattice spacing, 
so we will not be able 
to extrapolate the scalar mass 
to the continuum and to the infinite volume in 
the present work. The configurations were generated using DBW2 gauge action 
 \cite{rbc_dbw2}\footnote{DBW2 gauge action  is known to break the 
   reflection positivity of the transfer matrix \cite{Necco}, 
   which is the counterpart of the unitarity for Euclidean lattice theory.  
   Hence, there is a possibility that the negative  value of the scalar 
   correlator is partially caused by the complex eigenvalues of
   the transfer matrix. We concentrate our analysis at 
  large time separation $t \geq 6$,  where the negative 
   contribution to the two-point function is expected to disappear. 
  We perform the comparison between the full QCD and 
   the (partially) quenched QCD with $m_{val}<m_{sea}$ 
  using the  same DBW2 gauge action and find negative 
	scalar correlator only in case of the (partially) quenched QCD.  } 
with $\beta=0.80$ and Domain Wall fermion action \cite{dwf}  with $M_5=1.8$ and $L_s=12$ ($L_s$ is the extent in $5$th dimension) \cite{dyn_dwf}. The separate evolutions were performed for three different bare sea-quark masses $m_{sea}=0.02,~0.03,~0.04$, which correspond approximately to $M_\pi\!\sim\! 500\!-700\!~$MeV. The measurements of the correlators were performed on configurations separated by $50$ HMC trajectories.  Dynamical Domain Wall fermions have good chiral properties even at finite $L_s$ with the additive shift in
the mass due to the residual chiral symmetry breaking, $m_{res}$, being
approximately $0.0014$  \cite{dyn_dwf},  
much smaller than either the input sea or valence quark masses. The inverse lattice spacing was determined  from the $\rho$-meson mass and the preliminary result is  $a^{-1}\approx 1.7$ GeV \cite{dyn_dwf}. The current uncertainty of the lattice spacing  has a small effect on the scalar mass; the uncertainty of the scalar meson mass in the present work is dominated by the statistical errors of the scalar correlators.  

The scalar correlators were measured for the degenerate valence quark masses $m_1=m_2\equiv m_{val}$ in the range $m_{val}=0.01-0.05$. 
These valence quark masses 
 correspond approximately to $M_\pi\!\sim\! 380\!-\!770~$MeV (Fig. \ref{fig.pi}). We simulated the correlators with $\vec p=0$, point source and point sink via
$$\frac{1}{N_L^3}\sum_{\vec x,\vec x^\prime,\vec y}\langle 0|\bar q(\vec x,t)\Gamma q(\vec x^{\prime},t)~\bar q(\vec y,0)\Gamma q(\vec y,0)|0\rangle$$
  with $\Gamma=I$ on the lattices with un-fixed gauge.
 The average over the
configurations with un-fixed gauge gives the point-point correlator \cite{kuramashi}
\begin{equation}
\label{pp}
C_{pp}=\frac{1}{N_L^3}\sum_{\vec x,\vec y}\langle 0|\bar q(\vec x,t)\Gamma q(\vec x,t)~\bar q(\vec y,0)\Gamma q(\vec y,0)|0\rangle~.
\end{equation}
This method enabled us to calculate also the
 singlet scalar and singlet pseudoscalar correlators and to determine the hairpin insertion $m_0$. 

The summary of the scalar and pseudoscalar correlators analyzed in the present paper is given in the Table \ref{tab.data}.

\section{Analysis of dynamical correlators with $m_{val}=m_{sea}$}

\begin{figure}[htb!]
\begin{center}
\epsfig{file=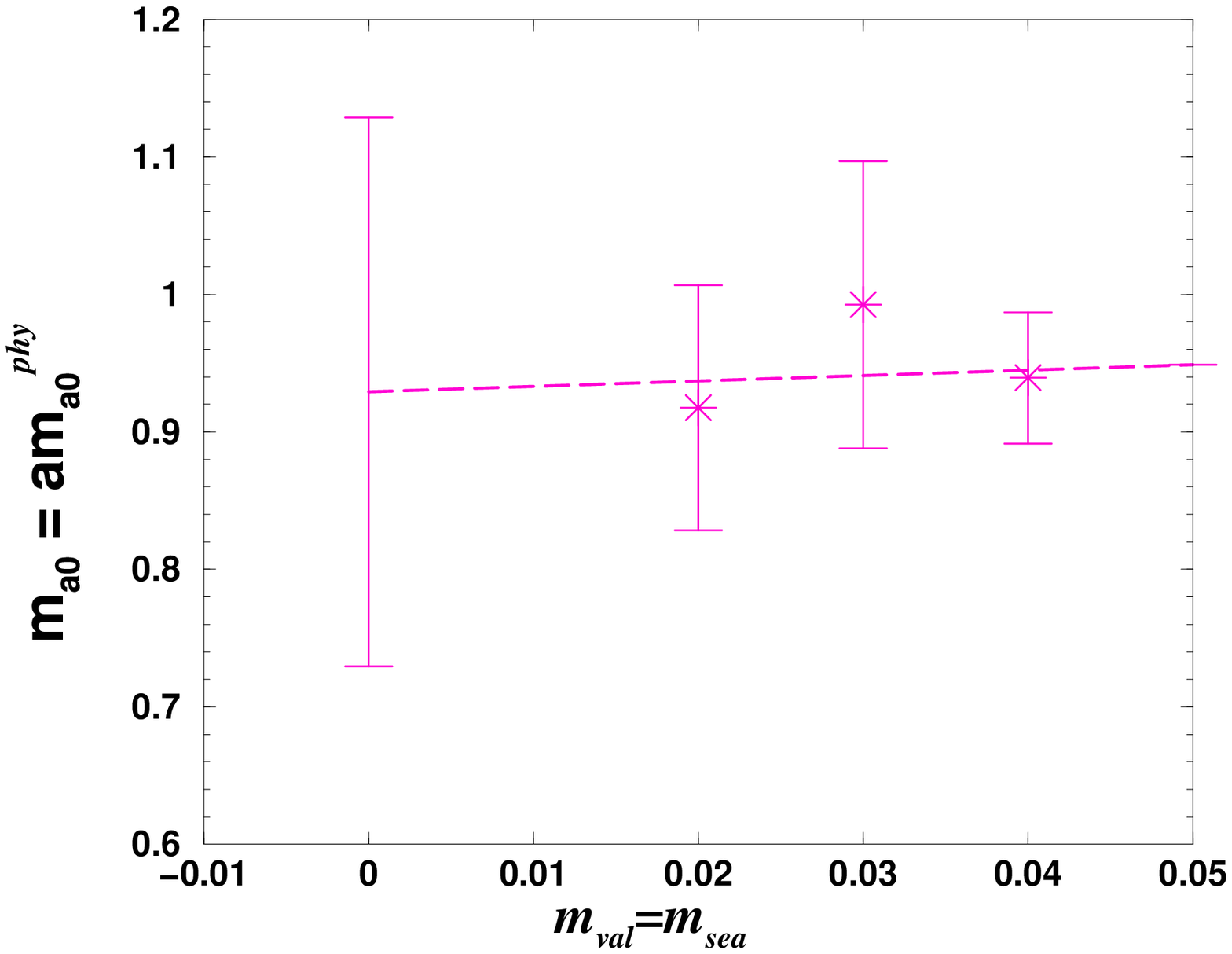,width=7.5cm}
$\quad$
\epsfig{file=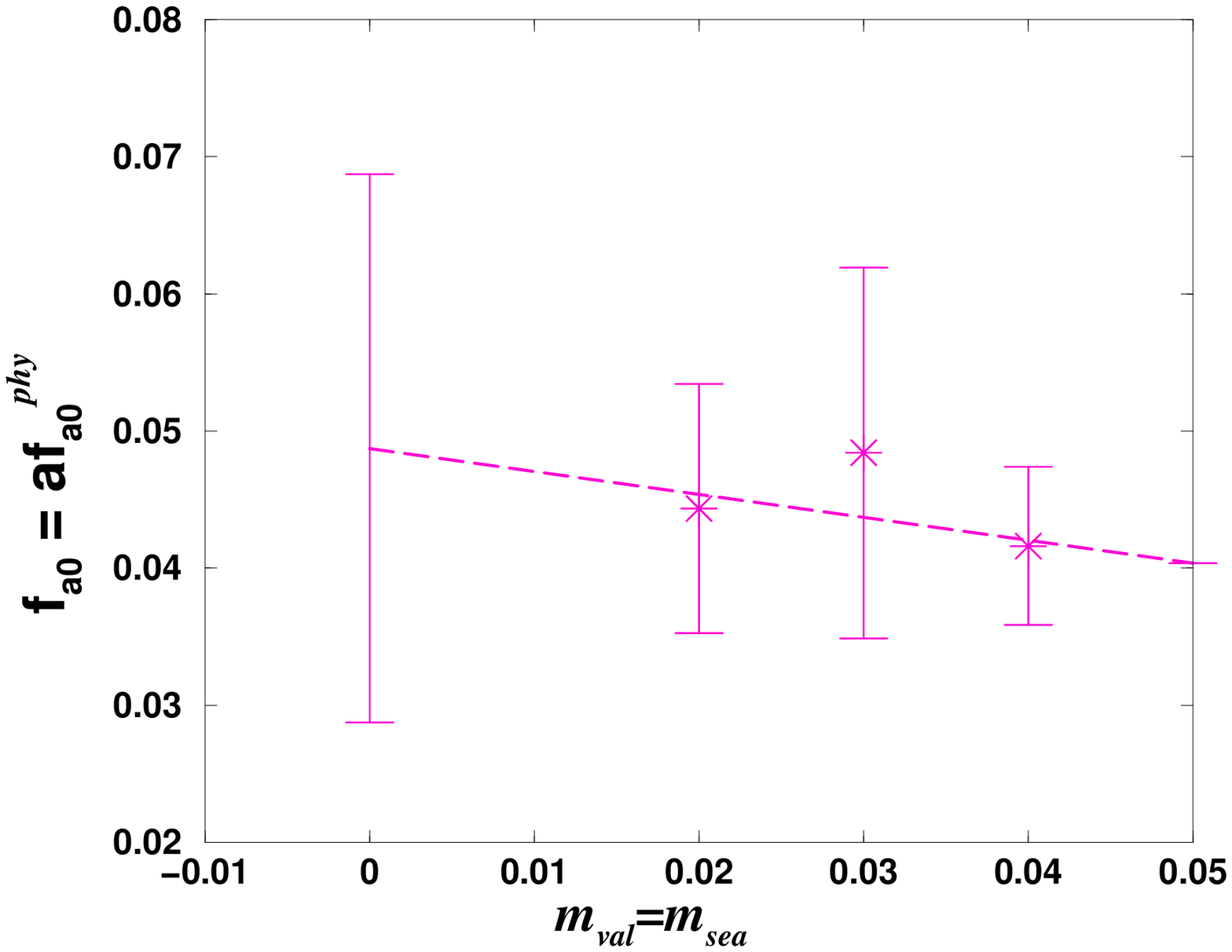,width=7.5cm}
\end{center}

\vspace{-0.8cm}

\caption{\small 
The asterisks represent the mass $m_{a0}$ and decay constant $f_{a0}$ 
 in the lattice units 
at the dynamical point $m_{val}=m_{sea}$, which are obtained from the exponential fit of the scalar correlators. 
The dashed line and the value at $m_q=0$ are obtained with 
 the linear fit.}\label{fig.dyn}
\end{figure}

The mass $m_{a0}$ and the unrenormalized 
decay constant $f_{a0}$ can be extracted from the
dynamical  scalar correlators using the exponential fit 
in the conventional way. Indeed, 
we will verify that the additional contribution from the exchange of two 
pseudoscalar fields in PQChPT (Fig. \ref{fig.bubble}b) 
exactly vanishes for 
$m_{val}=m_{sea}$, $N_{f}=2$  and $m_0\to\infty$ (\ref{Bfull}), 
so the simple exponential fit is well 
justified. The extracted masses and decay constants 
are shown in Fig. \ref{fig.dyn},  while Table \ref{tab.dyn} presents also the time ranges $t=t_{min}-t_{max}$ and $\chi^2$ of the fit\footnote{$t_{min}$ is taken throughout this work high enough so that there is no visible effect of the excited states and at the same time as low as possible in order to avoid large statistical errors. The choice of $t_{max}$ has negligible effect on the result and we take it at the time slice just before the signal is lost.}. The uncorrelated fits are used throughout this work and  the error-bars are obtained using the jack-knife method. The linear extrapolation to the chiral limit $m_{val}=m_{sea}\to 0$ gives results in lattice units\footnote{The difference between the chiral extrapolations $m_q\to 0$ and $m_q\to -m_{res}$ is negligible due to the smallness of $m_{res}\approx 0.0014$ \cite{dyn_dwf}.} 
\begin{equation}
\label{result_dyn}
m_{a0}=0.93(20)\ , \ f_{a0}=0.049(20)~, 
\end{equation}
where the jack-knife error-bars are calculated as described in Appendix B of \cite{extrapolation}.

The resulting errors are rather large, which motivates us to 
extract the mass also from the partially quenched data with 
$m_{val}\not =m_{sea}$. This forces us to understand the effect of partial 
quenching  in the following sections. The use of the Partially Quenched ChPT 
is crucial for this purpose since it enables us to subtract the 
significant partially quenched artifacts from the negative lattice correlators 
in case of $m_{val}<m_{sea}$.  
   
\section{Scalar correlator in partially quenched ChPT}

In this section we derive the non-singlet scalar correlator in the 
Partially Quenched ChPT (PQChPT) within the so-called $p$-expansion regime.  
We consider the theory with $N_{val}$ valence quarks 
$q_i$   (which can have different masses $m_i$) and $N_{f}$ degenerate 
sea quarks
$q_S$ of mass $m_{sea}$. The theory incorporates also $N_{val}$ valence ghost-quarks $\tilde q_{i}$ of mass $m_i$, which cancel the closed 
valence-quark loops.    PQChPT 
enables us to study of the partially quenched artifacts, which arise if 
the valence and the 
sea quark masses are not equal and if $N_{f}\not = N_{val}$. Our few lowest 
quark masses are low enough that $M_\pi^2/(4\pi f)^2\ll 1$, while 
they are still large enough that $M_\pi L\gg 1$ and we do not  enter 
$\epsilon$-regime on our lattice.

\begin{figure}[h]
\begin{center}
\epsfig{file=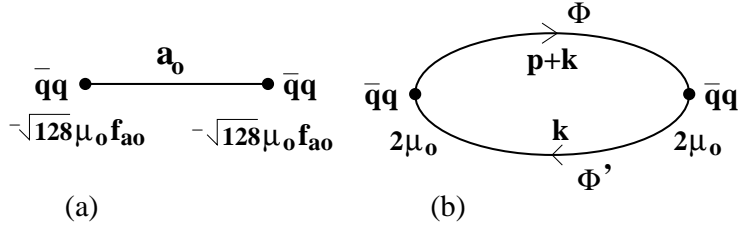,height=3cm}
\end{center}

\vspace{-0.8cm}

\caption{\small The contributions to the non-singlet scalar correlator in PQChPT: (a) The exchange of the scalar meson $a_0$; (b) The bubble diagram is responsible for the unphysical effect of partial quenching and represents the exchange of two pseudoscalar fields $\Phi\Phi^\prime$. The intermediate pseudoscalar 
 fields $\Phi$ and $\Phi^\prime$ 
can have the flavor structure $\Phi,\Phi^\prime\sim\bar q_iq_j,~\bar q_i\tilde q_{j},~\bar{\tilde {q_i}}q_{j},~\bar q_iq_S,~\bar q_Sq_i~$, where $q_{i,j}$ are the valence quarks and $q_S$ the sea quark.  }\label{fig.bubble}
\end{figure}

Non-physical contributions to the scalar correlator in 
PQChPT arise from the exchange of pseudoscalar fields 
between  $\bar q q$ source and $\bar q q$ sink. The leading 
contribution in the chiral expansion comes from the exchange of 
two pseudoscalar fields and is represented by the so-called 
{\it bubble  diagram} in Fig. \ref{fig.bubble}b. The two 
 pseudoscalar fields $\Phi$ and $\Phi^\prime$  can be mesons
$\Phi_{ij}\sim\bar q_iq_j$ and $\Phi_{iS}\sim \bar q_iq_S$ with Boson 
statistics, or  
mesons $\Phi_{i\tilde{j}}\sim \bar q_i\tilde q_{j}$ with fermionic 
statistics. 
We do not consider  the next-to-leading chiral 
corrections in PQChPT, which  are suppressed by ${\cal O}(M_\pi^2/(4\pi f)^2)$ 
in comparison to the bubble diagram\footnote{The NLO chiral 
corrections were taken into account for fully quenched scalar 
correlator by resummation  \cite{bardeen,bardeen2,sasa}.  
 They are more complicated in partially quenched theory
 because partially quenched theory implies several
 intermediate states $\Phi\Phi^\prime$ in Fig. \ref{fig.bubble}b, 
while quenched theory implies a single intermediate state
 $\pi\eta^\prime$.}. 
The  lattice correlators can be interpreted as a sum of the 
$a_0$-exchange at the tree level in Fig. \ref{fig.bubble}a and the 
 bubble  diagram in Fig. \ref{fig.bubble}b.
   
For our purpose, we need the strong 
interactions of pseudoscalar fields in PQChPT \cite{pqchpt} 
as well as the kinetic and the mass term for the $a_0$ 
field \footnote{Similar Lagrangian was used for the case of 
fully quenched ChPT in \cite{bardeen,bardeen2,sasa}. There $\mu_0$ is denoted 
by $\tfrac{1}{2}r_0$.  }
\begin{equation}
\label{lagrangian}
{\cal L}=\frac{f^2}{4}\str[\partial^\mu U \partial_\mu U^\dagger]+f^2\mu_0 \str[{\cal M}^\dagger U+U^\dagger {\cal M}]-\frac{m_0^2}{6}(\str \Phi)^2+\partial^\mu a_0\partial_\mu a_0^\dagger-m_{a0}^2 a_0 a_0^\dagger~
\end{equation}
 with the physical values $f\sim 95$ MeV and 
three-flavor $m_0\!\sim\! 600~$MeV$-1000~$MeV 
 \cite{m0}.  The field 
$U=\exp[\sqrt{2}i\Phi/f]$ incorporates the $SU(N_{val}+N_{f}|N_{val})_L\times
SU(N_{val}+N_{f}|N_{val})_R$ Goldstone field matrix  $\Phi$. 
The quark mass matrix ${\cal M}$ is 
$${\cal M}=diag(m_1,\dots,m_{N_{val}},\underbrace{m_{sea},\dots,m_{sea}}_{N_{f}},m_1,\dots,m_{N_{val}})$$
and the supertrace is defined as  $\str\! A=\sum_a\!\epsilon_a A_a$ with $\epsilon_a=1$ for quarks and $\epsilon_a=-1$ for ghost quarks.
The parameter
$\mu_0$ represents the slope of $M_\pi^2$ versus $m_q$ and the pseudoscalar masses  $M$ are given by  
\begin{equation}
M_{ab}^2=2\mu_0(m_a+m_b)\qquad,\quad  a,b=i,\tilde i,S\ , \quad i=1,..,N_{val}\ , \quad S=1,..,N_{f}~.
\end{equation} 
As is standard, we neglect the $\alpha(\partial \str \Phi)^2$ 
term in the Lagrangian since $\alpha$ seems to be small \cite{bardeen2,m0}.
 We also need the coupling of $a_0$ field  and the pseudoscalar 
 fields to the non-singlet scalar current\footnote{Similar current was used for the case of 
fully quenched ChPT in \cite{bardeen}. There $\mu_0$ is denoted 
by $\tfrac{1}{2}r_0$. }
\begin{equation}
\label{current}
\bar q_2 q_1\sim - \mu_0 f^2 (U+U^\dagger)_{12}-\sqrt{128}~\mu_0~f_{a0}~a_0
\end{equation}
where $f_{a0}$ plays the role  of the scalar meson decay constant with a given  normalization:
$$\langle 0 |\bar q_2 q_1|a_0\rangle=-~\sqrt{128}~\mu_0~f_{a0}~.$$

The point-point scalar correlator with external momentum $p=(\vec 0,E)$ 
\begin{equation}
\label{c_lat}
C^{lat}(t)\equiv \sum_{\vec x} \langle 0|\bar q_2(\vec x,t)q_1(\vec x,t)~\bar q_1(\vec 0,0)q_2(\vec 0,0)|0\rangle~
\end{equation}
is computed on the lattice and can be related to the prediction of PQChPT
\begin{align}
\label{c_pqchpt}
C^{PQChPT}(t)=F.T.\biggl[&\frac{128 \mu_0^2 f_{a0}^2}{p^2+m_{a0}^2}\biggr]~+~B(t)~.\\
&\ \ \  \downarrow\ {\rm continuum}\nonumber\\
\frac{128 \mu_0^2 f_{a0}^2}{2m_{a0}}&(e^{-m_{a0}t}\!+\!e^{-m_{a0}(N_T-t)})\nonumber
\end{align}
The first term is the conventional $a_0$-exchange, while the second 
term $B(t)=F.T.[B(p)]$ represents the contribution of the bubble diagram in 
 Fig. \ref{fig.bubble}b.
The lattice Euclidean momenta $p= 2\sin(p_E/2)$ and the 
discrete Fourier Transform F.T. on $p_E$ are used  
when we compare  PQChPT predictions (\ref{c_pqchpt}) 
with the lattice correlators\footnote{When we refer to conventional 
``exponential fit'', we extract $m_{a0}$ and $f_{a0}$ from the fit to the  
first term in (\ref{c_pqchpt}) using $p= 2\sin(p_E/2)$ and the discrete F.T.}. 
 The bubble diagram is calculated from the 
Lagrangian (\ref{lagrangian}) and the current (\ref{current}) in the  
Appendix A, giving
\begin{align}
\label{B}
&B(p)=4\mu_0^2 \! \sum_k\Biggl\{N_{f}~ \frac{1}{(k+p)^2+M_{1S}^2}~ \frac{1}{k^2+M_{2S}^2}\\
&\!\!\!\!\!\!\!-\frac{1}{N_{f}}~\frac{1}{(k+p)^2+M_{12}^2}~\frac{k^2+M_{SS}^2}{1+\frac{k^2+M_{SS}^2}{N_{f} ~m_0^2/3}}~\biggl[\frac{1}{(k^2+M_{11}^2)^2}+\frac{1}{(k^2+M_{22}^2)^2}+\frac{2}{(k^2+M_{11}^2)(k^2+M_{22}^2)}\biggr]\Biggl\}~.\nonumber
\end{align}
Here $k$ denotes the momenta in the loop (Fig. \ref{fig.bubble}b) 
 and the sum is performed over the allowed loop-momenta in the finite box of 
the lattice. We use the lattice Euclidean momenta $k= 2\sin(k_E/2)$ 
and $k+p= 2\sin((k_E+p_E)/2)$ when comparing  PQChPT prediction 
(\ref{c_pqchpt}) with the lattice correlators. The detailed expression 
for $C^{PQChPT}(t)$ used in case of our finite lattice is presented in Appendix B.

We note that the contribution of the bubble diagram (\ref{B}) has {\it no unknown parameters}, since one can fix the values of $M$, $\mu_0$ and $m_0$ from other considerations. We determine the pseudoscalar meson masses $M$ and $\mu_0=M^2/(4m_q)$ from pion correlators  on the same lattices. 
The hairpin insertion $m_0\!\sim\! 600~$MeV$-1000~$MeV 
(normalized for 3-flavors) has been determined 
from the $\eta^\prime$ 
 correlator in a number of references \cite{m0}, but the 
exact value of $m_0$ is not essential for the present work since the 
extracted scalar mass is almost independent of 
$m_0$ in the wide range $m_0\!=\![600~$MeV$,\infty]$  
 as will be demonstrated below.

\vspace{0.2cm}

In order to understand the effect of the bubble contribution 
on the lattice correlators, we derive the {\it asymptotic form of $B(t)$ 
at large $t$} for a correlator with $\vec p=0$ on a 
lattice with  $a\to 0$, $aN_T\to \infty$ and finite $aN_L$.
The asymptotic form for degenerate valence quarks with mass $m_{val}$ is 
\begin{align}
\label{B_as}
 &B(t)~\stackrel{t\to\infty}{\longrightarrow} ~\frac{2\mu_0^2}{N_L^3} \biggl[ 
  \frac{e^{-2M_{VS}~ t}}{M_{VS}^2} \frac{N_{f}}{2} ~+~
  \frac{e^{-(M_{VV}+M_{\eta'})t}} {M_{VV}~M_{\eta '}}  
          \frac{2N_{f} m_0^4} {9 (M_{\eta '}^2 -M_{VV}^2)^2}~ +\\
&\frac{e^{-2M_{VV}~ t}}{M_{VV}^2}
      \frac{m_0^2}{3M_{VV}^2}
  \biggl\{
-\frac{(M_{VV}^2\!-\!M_{SS}^2)^2+\tfrac{1}{3}N_{f}m_0^2(M_{VV}^2+M_{SS}^2)}{(M_{\eta '}^2-M_{VV}^2)^2}+\frac{M_{VV}^2-M_{SS}^2}{M_{\eta '}^2-M_{VV}^2} M_{VV}t \biggr\}  \biggr]~,\nonumber
\end{align} 
where $M_{\eta'}^2\equiv M_{SS}^2+\tfrac{1}{3}N_{f}m_0^2$ denotes
 $\eta'$ mass in a theory with $N_{f}$ flavors. 
We have assumed $M_{\eta'}>M_{VV}$ in derivation of (\ref{B_as}), which is 
satisfied for the pseudoscalar masses of physical interest.  
The asymptotic behavior is dominated by a pair of zero-momentum 
pseudoscalar fields with mass $M_{VV}\!=\!4\mu_0m_{val}~$,  a pair with mass
$M_{VS}\!=\!2\mu_0(m_{val}\!+\!m_{sea})$, or a pair with masses 
$M_{VV}$ and $M_{\eta '}$. The dominant 
contribution at large $t$ for  $m_{val}>m_{sea}$ is proportional to 
$e^{-2M_{VS} t}$ and has positive sign. The dominant 
contribution for  $m_{val}<m_{sea}$ is proportional to $t\cdot e^{-2M_{VV} t}$  and has negative sign given by $M_{VV}^2-M_{SS}^2$. The bubble contribution 
is inversely proportional to the spatial volume of the lattice, so the 
effect of the bubble contribution is much 
less important for larger lattices.
We  summarize these findings which apply\footnote{These findings apply as
 long as the condition 
$M_{\eta'}>M_{VV}$ is satisfied, which was assumed in the derivation of 
 (\ref{B_as}).} for any values of $N_{f}$ and $m_0$ as follows: 
\vspace{0.1cm}

 The scalar correlator receives a positive contribution $e^{-m_{a0}t}$
 from the exchange of $a_0$ meson and an additional bubble contribution 
from the exchange of two pseudoscalar fields $\Phi_1\Phi_2$. 
The bubble contribution is proportional to $e^{-(M_1+M_2)t}/N_L^3$ 
at large $t$ and it is positive for $m_{val}\geq m_{sea}$ and negative for 
$m_{val}<m_{sea}$. The scalar correlator with $m_{val}<m_{sea}$ 
has a negative sign at large $t$ if the bubble contribution 
dominates over the $a_0$-exchange.
The bubble contribution is particularly important 
on lattices with smaller spatial volume if $M_1+M_2<m_{a0}$. 
The magnitude of the bubble contribution  
is predicted by PQChPT (\ref{B}), 
which enables us to extract $m_{a0}$ and $f_{a0}$ 
by fitting the scalar correlators to PQChPT prediction 
(\ref{c_pqchpt}).  

\vspace{0.3cm}
 
We close this section by demonstrating that the analytical expressions for the bubble contribution (\ref{B},\ref{B_as}) reduce 
to the  known expressions in the fully quenched and the 
fully un-quenched limits:
\begin{itemize}
\item {\it The fully quenched ChPT} corresponds to the limit $m_{sea}\to\infty$ or equivalently 
to the limit $N_{f}\to 0$.  Eqs. (\ref{B}) and  (\ref{B_as}) 
reduce in this limit to

\vspace{-0.1cm}

\begin{align}
\label{BfullyQ}
B^{QChPT}(p)&=-16 \mu_0^2 \sum_k\frac{1}{(k+p)^2+M^2}~\frac{m_0^2/3}{(k^2+M^2)^2}~,\\
\label{BfullyQ_as}
B^{QChPT}(t)&\stackrel{t\to\infty}{\longrightarrow}~ -\frac{2\mu_0^2}{N_L^3}\frac{m_0^2}{3M^2}~(1+Mt)~\frac{e^{-2Mt}}{M^2}
\end{align}
with $M\equiv M_{VV}$ for degenerate valence quarks. 
 Expression (\ref{BfullyQ}) agrees with the fully quenched expressions  used in \cite{bardeen,bardeen2,sasa}.\footnote{The $m_0^2/3$ can be viewed as the hairpin insertion in the quenched theory with one valence flavor, while $\mu_0\equiv\tfrac{1}{2}r_0$ in \cite{bardeen,bardeen2,sasa}.} 
 
\item In the case of  {\it  full ChPT} with $SU(N_f)$ flavor symmetry  $m_1\!=\!m_2\!=\!m_{sea}$, the  expressions (\ref{B}) and (\ref{B_as}) reduce to
\begin{align}
\label{Bfull}
B^{ChPT}(p)&=4\mu_0^2~\frac{N_f^2-4}{N_f}~ \sum_k\frac{1}{(k+p)^2+M^2}~\frac{1}{k^2+M^2}\stackrel{N_f=2}{\longrightarrow} 0\qquad  {\rm for}\ m_0\to\infty~,\\
\label{Bfull_as}
B^{ChPT}(t)&\stackrel{t\to\infty}{\longrightarrow}~\frac{\mu_0^2}{N_L^3~N_f}
\biggl[ 4~\frac{e^{-(M+M_{\eta '})t}}{MM_{\eta '}}+
        (N_f^2-4)\frac{e^{-2Mt}}{M^2}\biggr]\qquad  {\rm for\ general}\ m_0~,
\end{align}
 where $M\equiv M_{VV}=M_{SS}$. 
This is in agreement with the conventional ChPT 
result. Bose symmetry and conservation of isospin allow
 only one intermediate state $\pi\eta^\prime$ in $N_f\!=\!2$ ChPT and 
 this intermediate state gives vanishing contribution
 in the $m_0\to \infty$ limit.
There are three intermediate states $\pi\eta$, $K\bar K$ and $\pi\eta'$ in 
$N_f\!=\!3$ ChPT.
\end{itemize}

\section{Analysis of lattice correlators using PQChPT }

Our  scalar correlators (\ref{c_lat}) are
 calculated in the partially quenched lattice QCD   for $N_{f}=2$ 
degenerate sea-quarks with mass $m_{sea}=0.02-0.04$ and degenerate valence quarks with mass $m_{val}=0.01-0.05$ (Table \ref{tab.data}). 
In this section we fit the lattice correlators 
to the one-loop prediction of the PQChPT given in Eq. (\ref{c_pqchpt}). The 
PQChPT prediction is the sum of the  $a_0$-exchange diagram  and the
bubble diagram (Fig. \ref{fig.bubble}). The magnitude of the bubble contribution (\ref{B}) is completely determined for a given choice of $m_0$ 
since we determine  the values of $M$ and $\mu_0$ from the pseudoscalar correlators on the same configurations:

\begin{figure}[htb!]
\begin{center}
\epsfig{file=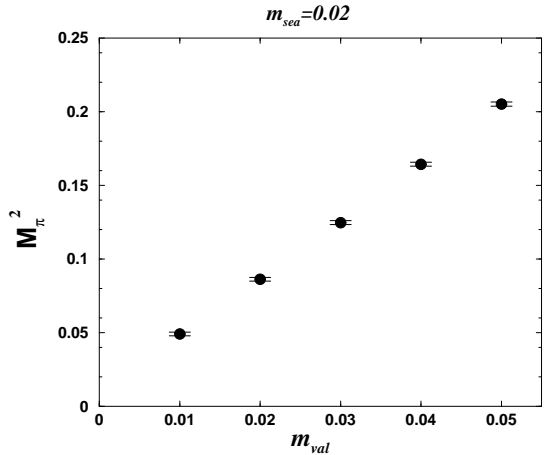,width=7cm}
\end{center}

\vspace{-1cm}

\caption{\small $M_\pi^2(m_{val})$ at $m_{sea}=0.02$ in lattice units. }\label{fig.pi}
\end{figure}
 
\begin{itemize}
\item The pseudoscalar masses are given by 
$M_{ab}^2=\tfrac{1}{2}(M_{aa}^2+M_{bb}^2)$, where $M_{aa}$ is determined 
from the pseudoscalar correlator with $m_{val}=m_{a}$ and the sea quark mass of interest. The $M_\pi$ from our lattice correlators are listed  in Table \ref{tab.pi} and shown in Fig. \ref{fig.pi}.
\item We fix $\mu_0=M_{val,val}^2/(4m_{val})$ for given 
$m_{val}$ and $m_{sea}$, where $m_{res}$ is neglected 
since it is much smaller than either $m_{val}$ or $m_{sea}$. 
 Here $m_{val}$ is the input bare mass of the valence quark. $M_{val,val}$ is the mass of the pion with two valence 
quarks of mass $m_{val}$ at $m_{sea}$ of interest. 
\end{itemize} 

The bubble diagram vanishes in the case of the dynamical theory ($m_{val}=m_{sea}$) with $N_{f}=2$ and $m_0\to\infty$ (\ref{Bfull}). In this case, the 
lattice correlator is interpreted solely by the $a_0$-exchange (\ref{c_pqchpt}), it has exponential time-dependence and the corresponding $m_{a0}$ was extracted in Section 3.
The bubble diagram is non-zero in general, so it has to be taken into account when the lattice correlators are fitted by equation (\ref{c_pqchpt}) in order to extract $m_{a0}$ and $f_{a0}$. 
The bubble contribution incorporates  the physical contributions from the 
exchange of two pseudoscalars and 
also the unphysical effects of partial quenching when $m_{val}\not = m_{sea}$.

We note that our scalar correlators and the quark masses, used to determine $\mu_0=M^2/(4m_q)$, are not renormalized. We would like to emphasize that this does not prevent us from extracting the physical mass  $m_{a0}$.  This can be seen by rearranging Eqs. (\ref{c_lat}) and (\ref{c_pqchpt}):
\begin{equation}
\label{renorm}
m_q^2~{\sum_{\vec x}} \langle 0|\bar q_2q_1(\vec x,t)~\bar q_1q_2(\vec 0,0)|0\rangle = coef.~\biggl(e^{-m_{a0}t}+e^{-m_{a0}(N_T-t)}\biggr)+F.T\biggl[\frac{M_\pi^4}{16}\frac{B(p)}{\mu_0^2}\biggr]~.
\end{equation}
The product of quark mass and the scalar current $m_q\bar q q$ is invariant under renormalization. 
 The second term on the RHS of (\ref{renorm}) depends only on the 
hadron masses $M_\pi$ and $m_0$, which are also invariant under  
renormalization. The scalar mass $m_{a0}$ can be therefore extracted from the first term on RHS without ambiguity from the renormalization.

\subsection{Analysis of dynamical correlators taking into account $\pi\eta^\prime$ intermediate state}

\begin{figure}[htb!]
\begin{center}
\epsfig{file=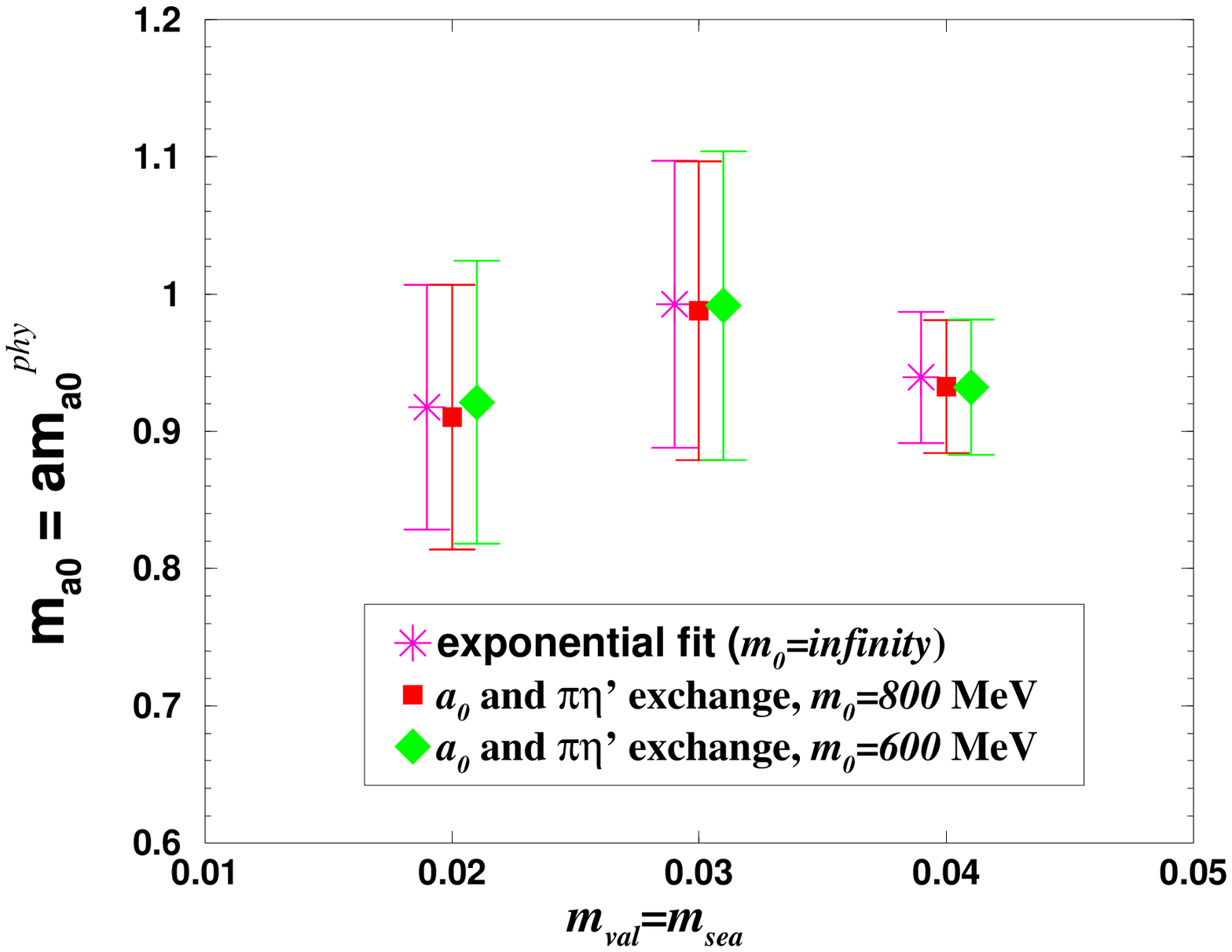,width=7.5cm}
$\quad$
\epsfig{file=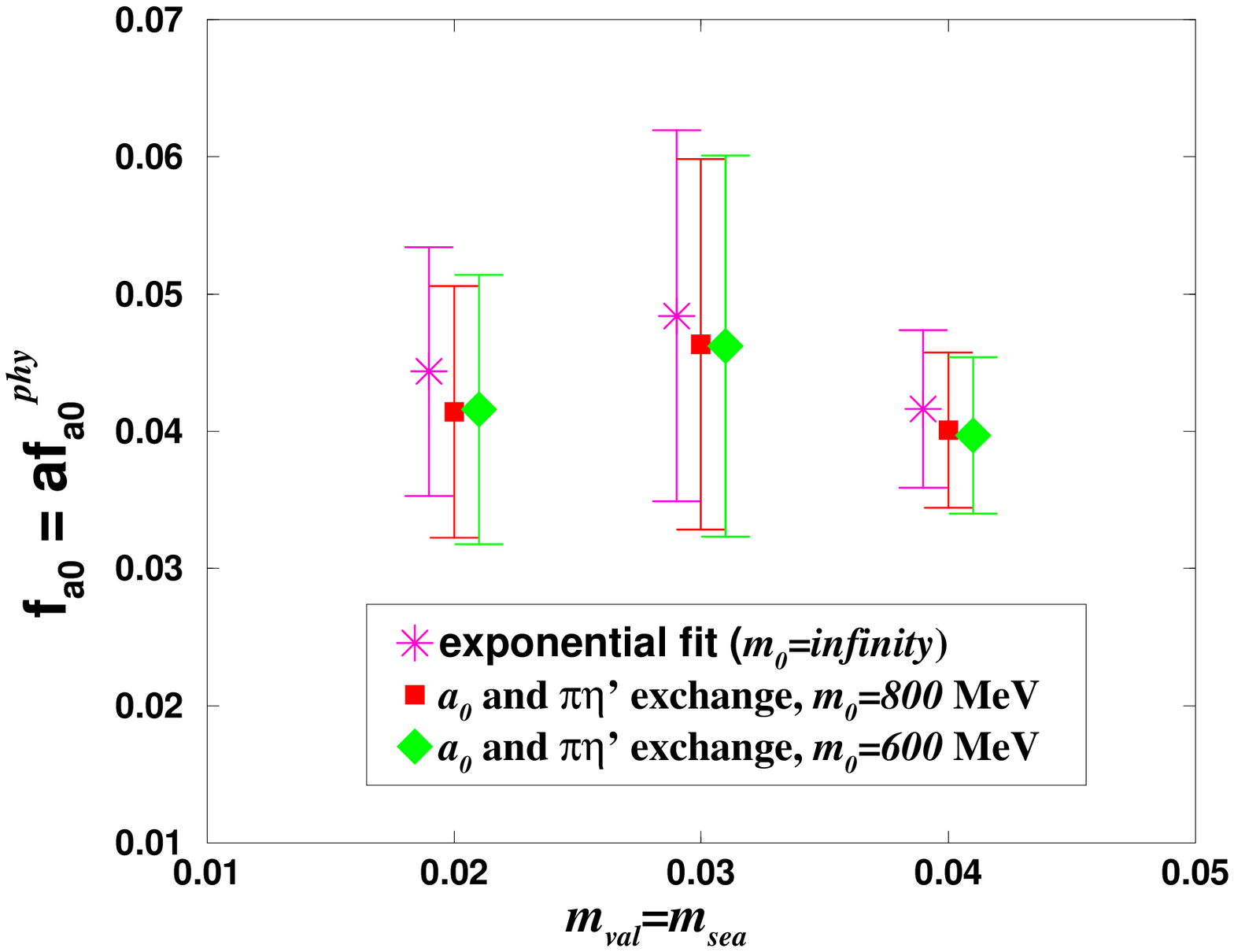,width=7.5cm}
\end{center}

\vspace{-0.8cm}

\caption{\small 
The $m_{a0}$ and $f_{a0}$ in lattice units  
obtained from the fit of the dynamical scalar correlators 
with the PQChPT prediction (\ref{c_pqchpt}) for various $m_0$.  
The PQChPT prediction for two-flavor dynamical correlators 
incorporates $a_0$ and $\pi\eta^\prime$ exchange. 
The asterisks represent the fit for $m_0\to\infty$,
when the contribution from $\pi\eta^\prime$ state vanishes 
and the fit reduces to the standard exponential fit  
(asterisks are the same as asterisks in Fig. \ref{fig.dyn}).
 The different data points are slightly 
shifted from $m_{q}=0.02,0.03,0.04$ in horizontal 
direction for clarity. }\label{fig.dyn_bubble}
\end{figure}

The two-flavor dynamical scalar correlator receives 
a contribution from the exchange of $a_0$ and from the 
exchange of $\pi\eta^\prime$.
 The contribution from $\pi\eta^\prime$ state vanishes in the limit 
 $m_0\to\infty$ and
the fit reduces to the standard exponential fit used in section 3.
The contribution of $\pi\eta'$ intermediate state at finite $m_0$ 
is given by the bubble contribution  
(\ref{B}) with $m_{val}=m_{sea}$ and $N_{f}=2$. 
The fit of the dynamical correlators to PQChPT prediction 
(\ref{c_pqchpt}) at various $m_0$  gives $m_{a0}$ and $f_{a0}$ in 
Fig. \ref{fig.dyn_bubble}. 
The results are almost independent of $m_0$ 
and agree with the result of the conventional exponential fit  
in section 3. For this reason we will refer to section 3 for our  
dynamical results in the Conclusions. 

The extracted $m_{a0}$ and $f_{a0}$ are almost insensitive to the 
presence of the $\pi\eta^\prime$ state since the contribution of
this state is at least an order of magnitude smaller than the 
dynamical lattice correlators in the fitted time-range for $m_0\geq 600~$MeV. 
Our dynamical correlators are     
dominated by the $a_0$ exchange although 
the mass $M_\pi\!+\!M_{\eta'}=M_\pi\!+\!(M_\pi^2+\tfrac{2}{3}m_0^2)^{1/2}$ is comparable to  the mass $m_{a0}$ in Table \ref{tab.dyn}. This can be attributed to our spacial volume 
$16^3$ which is large enough to suppress the contribution $e^{-(M_\pi+M_{\eta '})t}/N_L^3$  of the $\pi\eta^\prime$ exchange in Eq. (\ref{Bfull_as}).

\subsection{Analysis of  the  partially quenched correlators with $m_{val}\not =m_{sea}$ }

Our analysis of the partially quenched correlators is based on the correlators with $m_{sea}=0.02$ since a range of $m_{val}=0.01-0.05$ is available only 
in this case (see Table \ref{tab.data}). 

The lattice correlators (\ref{pp}) 
 and the PQChPT bubble contribution $B(t)$ (\ref{c_pqchpt},\ref{B})  are compared in Fig. \ref{fig.2} for $m_{sea}=0.02$ and various $m_{val}$.  
The PQChPT predictions in Figs. \ref{fig.2}b and 
\ref{fig.2}c show just the bubble 
contribution without the $a_0$-exchange contribution in order to 
indicate the importance of the bubble contribution for various $m_{val}$ 
at fixed $m_{sea}$. Fig. \ref{fig.2} exhibits good qualitative agreement 
between
 the lattice correlators and the PQChPT predictions at various $m_{val}$:

\vspace{-0.1cm}

\begin{figure}[htb!]
\begin{center}
\epsfig{file=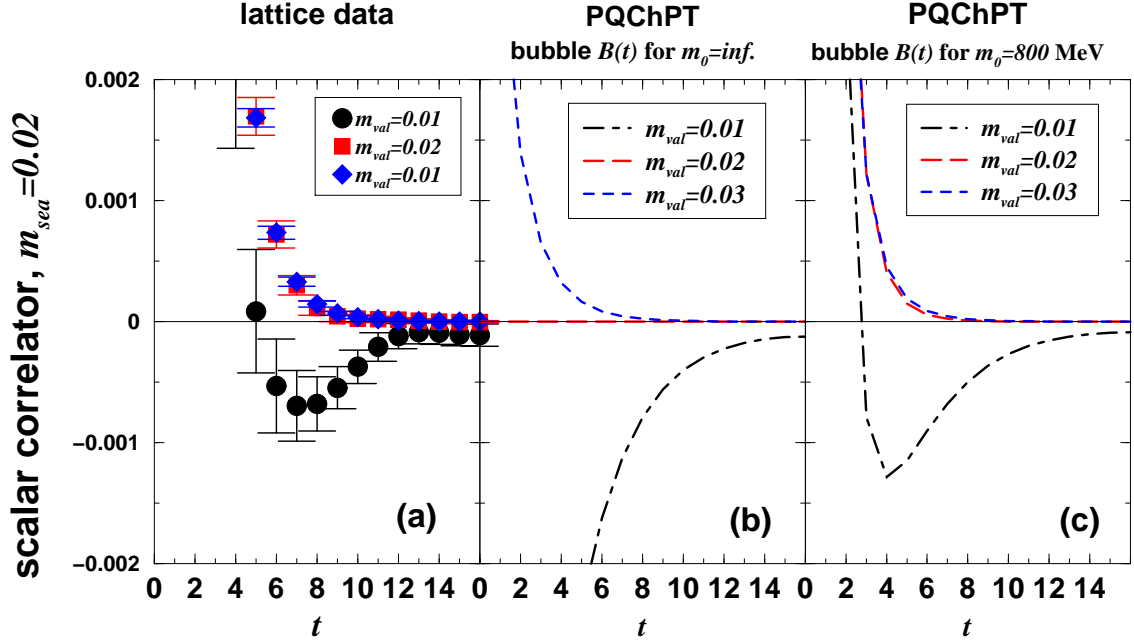,width=15cm}
\end{center}

\vspace{-1cm}

\caption{\small The scalar correlator for $m_{sea}=0.02$ and various $m_{val}$:  the lattice data (a) and the bubble contribution $B(t)$ as predicted by PQChPT for $m_0\to \infty$ (b) and  $m_0\to 800$ MeV (c). The PQChPT prediction (\ref{c_pqchpt}) in this figure represents just $B(t)$ and does not contain the contribution from the $a_0$-exchange. }\label{fig.2}
\end{figure}

\begin{itemize}

\vspace{-0.2cm}

\item The lattice correlator and the bubble contribution are both negative and large for $m_{val}\!<\!m_{sea}$ due to a striking unphysical effect of partial quenching. The bubble contribution is large and negative for $m_{val}=0.01$ since it falls like $-t\cdot e^{-2M_{VV}t}$ at large $t$ (\ref{B_as}), where  the corresponding pion mass $M_{VV}=0.222(3)$ is small (Table \ref{tab.pi}).

\vspace{-0.2cm} 

\item The dynamical lattice correlator with $m_{val}\!=\!m_{sea}$ is positive. The bubble contribution describes the exchange of physical $\pi\eta^\prime$ 
and vanishes in the limit $m_0\to \infty$ (\ref{Bfull}).

\vspace{-0.2cm}

\item The lattice correlator at $m_{val}>m_{sea}$ is positive and receives a positive and rather small contribution from the bubble, which is less and less important for larger $m_{val}$. The bubble contribution is positive and relatively small since it falls as a linear combination of $+e^{-2M_{VS}t}$ and $+t\cdot e^{-2M_{VV}t}$ at large $t$ (\ref{B_as}), where the pseudoscalar masses $M_{VS}$ and $M_{VV}$ 
are relatively large for $m_{val}\geq 0.03$ and $m_{sea}=0.02$.  
\end{itemize}

\vspace{-0.1cm}

While we do not include it  in the
extraction of our final results, we have a limited set of partially quenched
data for the $m_{sea}=0.03$ evolution, and we have checked that 
the lattice correlators change sign at $m_{val}=m_{sea}$ 
 also in this case. This can be seen in Fig. \ref{fig.1}b.

\vspace{0.1cm}

\begin{figure}[htb!]
\begin{center}
\epsfig{file=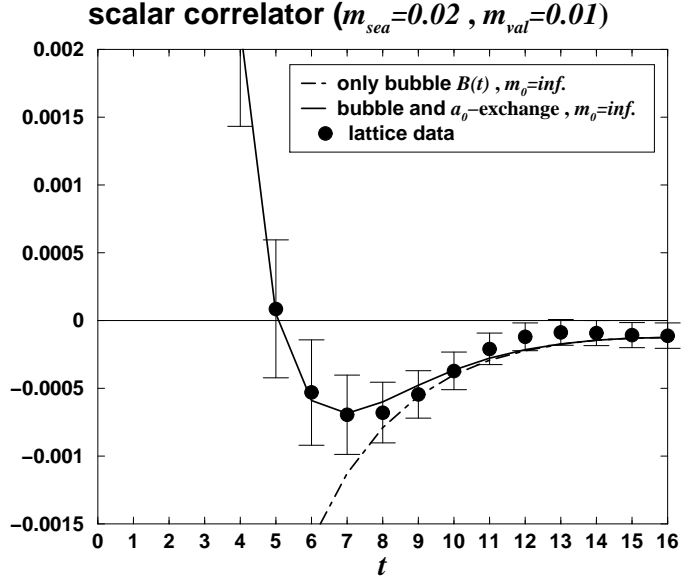,width=9cm}
\end{center}

\vspace{-1cm}

\caption{\small The data and PQChPT predictions for the scalar correlator at $m_{sea}=0.02$, $m_{val}=0.01$ and $m_0=\infty$. }\label{fig.3}
\end{figure}

\begin{figure}[htb!]
\begin{center}
\epsfig{file=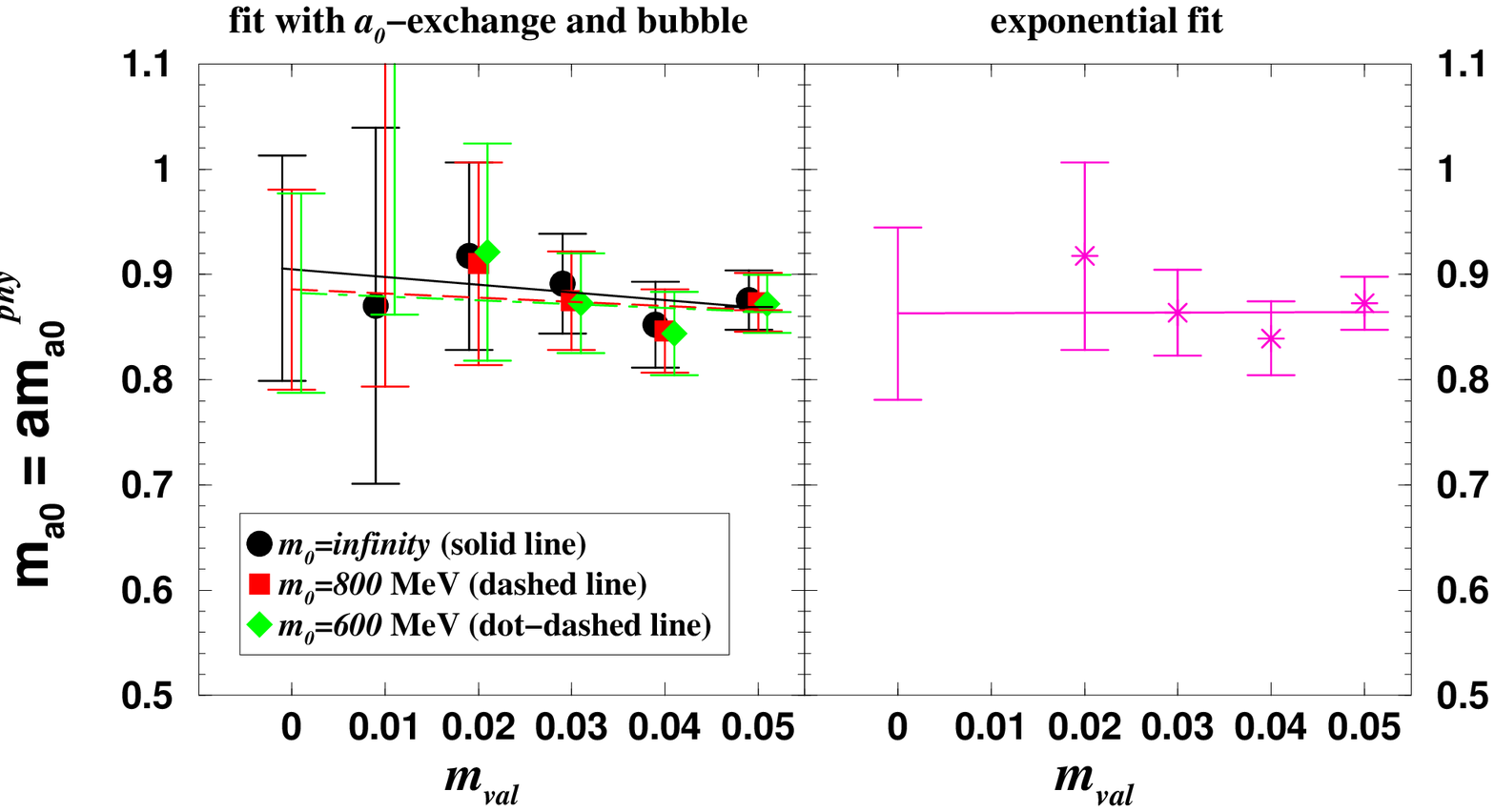,width=15cm}
\epsfig{file=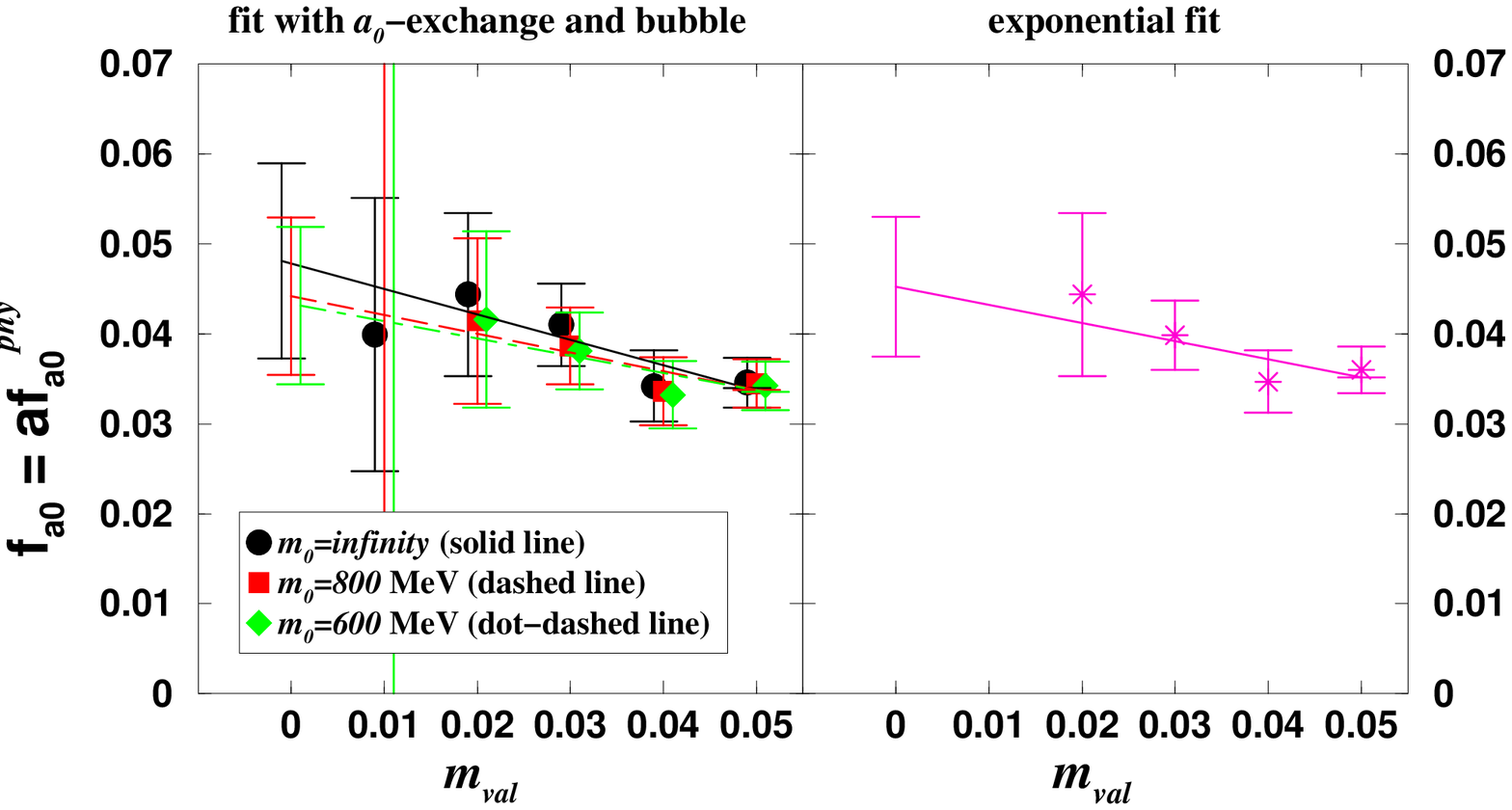,width=15cm}
\end{center}

\vspace{-0.9cm}

\caption{\small 
The $m_{a0}$ and $f_{a0}$ in lattice units 
obtained from the fit of the 
scalar correlators at $m_{sea}=0.02$ 
with the PQChPT prediction (\ref{c_pqchpt}). The left figures represent the 
fit results, when the bubble contribution is taken into account and 
$m_0$ is varied. The right figures represent the conventional exponential fit  $e^{-m_{a0}t}+e^{-m_{a0}(N_T-t)}$, which is obtained under the assumption that the bubble contribution vanishes;  the correlator with $m_{val}=0.01$ and $m_{sea}=0.02$ is negative and can not be described by $e^{-m_{a0}t}+e^{-m_{a0}(N_T-t)}$. The different data points are slightly 
shifted in horizontal direction for clarity. }\label{fig.4}
\end{figure}

Let us have a closer look at the case of $m_{val}=0.01$ and $m_{sea}=0.02$ in Fig. \ref{fig.3}, where the effect of partial quenching is most striking. The bubble contribution (dot-dashed) is in good quantitative agreement with the data for $t\geq 8$, where the $a_0$-exchange fades exponentially. The exchange of the $a_0$ scalar meson is dominant at smaller $t$. The fit of the data to the PQChPT prediction (\ref{c_pqchpt}) gives $m_{a0}\!=\!0.87(17)$ and $f_{a0}\!=\!0.040(15)$ at $m_0\to \infty$.  The PQCHPT prediction with this choice of 
$m_{a0}$ and $f_{a0}$ is given by the solid line  in Fig. \ref{fig.3} 
 and describes the lattice correlator well.   
Our one-loop analytical formula therefore  
correctly determines the sign and
approximate size of the effects when the valence quark mass is lower
than the sea quark mass. This gives us confidence in the veracity of
applying this formula to the larger valence quark masses, where the loop
effects are smaller.

All scalar correlators for $m_{sea}=0.02$ and various $m_{val}$ are fitted by the PQChPT prediction (\ref{c_pqchpt}),  and the resulting 
 $m_{a0}$ and $f_{a0}$ are given in Fig. \ref{fig.4} and Table \ref{tab.4}. 
Figures on the left represent the 
 fit, which incorporates both the bubble and the $a_0$-exchange 
contributions. We find that $m_{a0}$ and $f_{a0}$ at $m_{val}\geq 0.02$ 
depend very slightly on the hairpin insertion $m_0$ in the range 
$m_0=[600~$MeV$,\infty]$. In the case of $m_{val}=0.01$, the central values of $m_{a0}$ and $f_{a0}$ depend significantly on $m_0$, but they are all consistent for $m_0$ in the range $[600~$MeV$,\infty]$ within the large error-bars\footnote{ Large error-bars on $m_{a0}$ and $f_{a0}$ at $m_{val}=0.01$ arise since 
they are obtained from the fit to the 
difference of the lattice correlator and the bubble contribution. The lattice correlator and the bubble contribution are negative and large for $m_{val}=0.01$, so their difference is small and has relatively large error-bar.}.  
The result of the linear extrapolation from $m_{val}=0.01-0.05$ to the chiral limit $m_{val}\to 0$ is practically independent of whether the $m_{val}=0.01$ data is taken into account due to the large error-bars at $m_{val}=0.01$ with current statistics. The linear extrapolation from   $m_{val}=0.01-0.05$ gives
\begin{align}
m_{a0}&=0.90(11)~,\ f_{a0}=0.048(11)\ \ {\rm for}\ m_0\to \infty \nonumber\\ 
m_{a0}&=0.89(9)~, \ f_{a0}=0.044(9)\ \ {\rm for}\ m_0= 800~{\rm MeV}\\
m_{a0}&=0.88(9)~, \ f_{a0}=0.043(9)\ \ {\rm for}\ m_0=600~{\rm MeV\ ,}\nonumber
\end{align}
which are consistent for $m_0=[600~$MeV$,\infty]$. 
So the chiral extrapolation $m_{val}\to 0$ at fixed $m_{sea}=0.02$ leads to 
the mass in the lattice units 
\begin{equation}
 \label{result_pq}
m_{a0}=0.89(11)~,
\end{equation}
where the error reflects the statistical error of the data
and the variation of the bubble contribution for $m_0$ in the range  
 $m_0=[600~$MeV$,\infty]$.

The conventional exponential fit of the scalar correlators
for  $m_{val}\geq m_{sea}\!=\!0.02$ 
 gives $m_{a0}$ and $f_{a0}$ in Fig. \ref{fig.4} on the right. 
The bubble contribution in Eq. (\ref{c_pqchpt})  is taken to be zero in 
this case. The exponential fit obviously does not work for the correlator at 
$m_{val}=0.01$, where the intriguing  partially quenched artifact has to be 
incorporated through the bubble contribution.  
 However, it gives reasonable 
$m_{a0}$ and $f_{a0}$  for $m_{val}\geq m_{sea}$:  
the results from the exponential fit are consistent  with the 
results from the fit to  Eq. (\ref{c_pqchpt}) 
since the bubble contribution is zero or relatively small for 
$m_{val}\geq m_{sea}$.

\section{Non-singlet scalar meson mass}

In this section we collect our main results on the scalar meson mass.  

The chiral extrapolation 
$m_q\to 0$ of the two-flavor {\it dynamical}  
 data points $m_{val}\!=\! m_{sea}\!=\!m_q$ gives (Eq. \ref{result_dyn})
\begin{equation} 
\label{res_dyn}
m_{a0}=0.93\pm 0.20\quad {\rm or}\quad  m_{a0}^{phy}=1.58\pm 0.34~{\rm GeV}~,
\end{equation}
where only the statistical error is given. 
 The number in GeV is obtained 
using the preliminary result for the scale 
$a^{-1}\approx\! 1.7~$GeV \cite{dyn_dwf}. 
 
We extracted also the scalar meson masses from the {\it partially quenched}
 correlators with $m_{val}\!\not =\! m_{sea}$. 
The chiral extrapolation  $m_{val}\to 0$ at fixed $m_{sea}=0.02$ 
 leads to 
\begin{equation}
\label{res_pq}
m_{a0}=0.89\pm 0.11  \quad {\rm or}\quad  m_{a0}^{phy}=1.51\pm 0.19~{\rm GeV}~
\end{equation}
and we expect that the dependence on the sea quark mass is small.
Here the error reflects the statistical error of the data
and the variation of the bubble contribution for $m_0$ in the range  
 $m_0=[600~$MeV$,\infty]$ (see  Eq. \ref{result_pq} and Fig. \ref{fig.4}). 

The chiral limits of $m_{a0}$ in the dynamical case and in the partially 
quenched case are consistent. 
Note that the error is  smaller in the partially quenched case, 
where the application of the Partially Quenched ChPT was crucial.  
The mass of the simulated $q\bar q$ state is somewhat larger than the mass of  
$a_0(980)$ and it is closer to the mass of $a_0(1450)$. We note that 
our result is consistent with the fully quenched results of Refs. \cite{bardeen,bardeen2,sasa} within the present accuracy\footnote{The fully quenched results of \cite{bardeen,bardeen2,sasa} are presented in the Introduction and are consistent with (\ref{res_dyn},\ref{res_pq}) if the effect of quenching 
 is incorporated at the leading order 
in the chiral expansion (one bubble). The quenched $m_{a0}$ is somewhat lower if quenching effect is incorporated at the next-to-leading order \cite{sasa}.}. 

\vspace{0.2cm}

Finally we comment on the scalar iso-doublet mesons $s\bar u$ and $s\bar d$, since their relation to the controversial resonance $\kappa$ is still an open question.  We are not able to make a reliable estimate for the mass of the $s\bar u$ and $s\bar d$ scalar mesons, since we did not simulate non-degenerate valence quarks. We get a rough estimate  by extrapolating 
the mass obtained from the dynamical correlators 
to $\tfrac{1}{2}(m_s+m_{u,d})$. The resulting mass $m_{\kappa}\sim 0.92(9)$ or $m_\kappa^{phy}\sim 1.6\pm 0.2~$GeV seems higher than the mass of the reported 
experimental candidate ( $\sim 800~$MeV \cite{kappa}) although the interpretation  of this observation as $\kappa$ is controversial.

\section{Conclusions}

We presented a  lattice study of the lightest  scalar $q\bar q$ state with
non-singlet flavor structure ($a_0$ meson).
Good chiral properties of the Domain Wall Fermions
are important for the connected scalar correlator since this is
the first step toward a controlled investigation of the
scalar spectrum, in particular, the
$\sigma$ particle, which is intimately related to the chiral
symmetry breaking. Two degenerate  sea-quarks
were simulated with masses corresponding to $M_\pi\!\sim\!500-700~$MeV.
The simulations were done at fixed lattice spacing and one size of the
volume.

The value of scalar mass $m_{a0}=0.93\pm 0.20$ in the lattice units
was extracted in the conventional way from the dynamical correlators
($m_{val}\!
=\!m_{sea}$) and the resulting error  is rather large. The corresponding
physical mass $m_{a0}^{phy}=1.58\pm 0.34~$GeV was obtained using the
preliminary result for the scale $a^{-1}\!\approx\!  1.7~$GeV.

We analyzed also the partially quenched correlators with
$m_{val}\!\not= \! m_{sea}$. They exhibit striking effect of
partial quenching
 since they are positive for $m_{val}\geq m_{sea}$ and negative for
 $m_{val}<m_{sea}$ (Fig. \ref{fig.1}). In order to understand this
effect of partial quenching, we
derived the scalar correlator within the Partially Quenched ChPT.
The leading unphysical contribution comes from the
exchange of two pseudoscalar fields and has no unknown parameters.
We have shown that this contribution is positive for
 $m_{val}\geq m_{sea}$, it is negative for $m_{val}< m_{sea}$ and
it is inversely proportional to spatial-volume at large
time-separations.
The physical contribution to the scalar correlator
is due to the exchange of the scalar meson
$a_0$ and has conventional  exponential time-dependence.
We find that the sum of these two contributions describes our
partially quenched lattice correlators very well.  We extract the mass
$m_{a0}=0.89\pm 0.11$ or $m_{a0}^{phy}=1.51\pm 0.19~$GeV from partially 
quenched correlators,
which is consistent with
the mass extracted from our dynamical correlators.

Our current simulation of the $q\bar q$ state on the lattice
seems to indicate
that this state is somewhat heavier than
the observed resonance $a_0(980)$,
and it is closer to the observed resonance $a_0(1450)$;
however, given the size of our errors this is not conclusive.
We  must also emphasize that we have
only two dynamical flavors, our lattice volume is not
very large and also our quark masses are quite heavy. Besides,
continuum limit has not been taken as we have data at only one
lattice spacing, so
the exploratory nature of our study needs to be kept in mind.

We also note that
 the $N_f=2$ theory is  likely to have interesting differences
from QCD ($N_f = 2 + 1$). Recall that the observed resonances  
$a_0(980)$ and $a_0(1450)$  decay to $\eta\pi$ and
$K \bar K$. Bose statistics and isospin
conservation restrict $a_o^{N_f=2} \not \rightarrow \pi + \pi$ for $N_f=2$,
though $a_0^{N_f=2} \rightarrow \eta'^{N_f=2} + \pi$ would be possible 
if kinematics allows it. 
Additional intricacy could be also caused by the 
presence of a large 4-quark component in these channels.  
 Thus the approach to the
chiral limit may well exhibit a more involved dependence of the scalar
mass on the quark mass than our data (see Fig. 2) indicates
with relatively heavy quarks.
These issues will need to be addressed in future works with more
computing resources.

\vspace{1cm}

{\bf  \large Acknowledgments}

\vspace{0.3cm}

The dynamical Domain Wall Fermion configurations were generated by the 
RBC Collaboration and were essential for the lattice results in 
the present paper. 
It is a pleasure to thank all the members of the RBC Collaboration, 
in particular Tom Blum, Yasumichi Aoki, Norman Christ, Bob Mawhinney, 
Shigemi Ohta and June Noaki. We also 
thank RIKEN, Brookhaven National Laboratory and U.S. Department of Energy 
for providing the facilities essential for the completion of this work. All 
computations were carried out on the QCDSP supercomputers at the RIKEN BNL 
Research Center and at Columbia University. 
The research of A.S. was supported in part by the USDOE
contract number DE-AC02-98CH10886. K.O. was supported in part
by D.O.E. grant DFFC02-94ER40818.


\newpage

\vspace{0.5cm}

\appendix
\refstepcounter{section}
\section*{Appendix A: Calculation of the bubble diagram}\label{app}

In this appendix we derive the result $B(p)$ (Eq. \ref{B}) for the bubble diagram in Fig. \ref{fig.bubble}b. The coupling of the non-singlet scalar current and 
two pseudoscalar fields $\Phi$ is given by Eq. (\ref{current})
\begin{equation}
\bar q_2 q_1=2\mu_0 (\Phi^2)_{12}~.
\end{equation}
The scalar correlator receives the following contribution from the 
exchange of the two pseudoscalar fields shown in Fig. \ref{fig.bubble}b
\begin{equation}
B=\langle 0|\bar q_2 q_1~\bar q_1 q_2|0\rangle_{Fig. \ref{fig.bubble}b}=
4 \mu_0^2 \langle 0|\Phi_{1a}\Phi_{a2}~ \Phi_{2b}\Phi_{b1}|0\rangle~,
\end{equation}
where indices $a$ and $b$ are summed over all quarks and ghost-quarks of the
partially quenched theory: $a,b=i,\tilde i,S$ with $i=1,..,N_{val}$ and $S=1,..,N_{f}$. The non-zero Wick contractions relevant to the diagram on the Fig. 
\ref{fig.bubble}b are
\begin{align}
\label{derivation}
B=4 \mu_0^2\biggl( &\langle\Phi_{1S}|\Phi_{S1}\rangle\langle\Phi_{S2}|\Phi_{2S}\rangle+\langle\Phi_{1i}|\Phi_{i 1}\rangle \langle\Phi_{i 2}|\Phi_{2 i}\rangle+\langle\Phi_{1\tilde i}|\Phi_{\tilde i 1}\rangle \langle\Phi_{\tilde i 2}|\Phi_{2\tilde i}\rangle\nonumber\\
+&\langle\Phi_{11}|\Phi_{22}\rangle\langle\Phi_{12}|\Phi_{21}\rangle+\langle\Phi_{22}|\Phi_{11}\rangle\langle\Phi_{12}|\Phi_{21}\rangle\biggr)~.
\end{align}
The propagators for the pseudoscalar fields follow from the Lagrangian (\ref{lagrangian}).   The propagator for the flavor non-diagonal mesons ($a\not =b$) in Minkowski space is
\begin{equation}
\label{non-diagonal}
\langle\Phi_{ab}|\Phi_{ba}\rangle=i\frac{\delta_{ab} \epsilon_a}{p^2-M_{aa}^2}~,
\end{equation}
while for the diagonal mesons the propagator is \cite{pqchpt}
 \begin{equation}
\label{diagonal}
\langle\Phi_{aa}|\Phi_{bb}\rangle=i\biggl[\frac{\delta_{ab} \epsilon_a}{p^2-M_{aa}^2}-\frac{1}{N_{f}}\frac{p^2-M_{SS}^2}{(p^2-M_{aa}^2)(p^2-M_{bb}^2)}~\frac{1}{1-\frac{p^2-M_{SS}^2}{N_{f}~ m_0^2/3}}~\biggr]~.
\end{equation}   
The analytical expression (Eq. \ref{B}) for the bubble diagram in Fig. \ref{fig.bubble} is obtained by inserting the propagators (\ref{diagonal}) and (\ref{non-diagonal}) to the expression (\ref{derivation}) and by performing the Wick rotation to the Euclidean space.

\section*{Appendix B: PQChPT correlator for a finite lattice }  

The PQChPT prediction for the scalar  correlator 
$C^{PQChPT}(t)$ (\ref{c_pqchpt}) relevant for a finite 
lattice of the volume $N_L^3N_T$ is
\begin{align}
&C^{PQChPT}(t)=\frac{1}{N_T}\sum_{m_4=-N_T/2}^{N_T/2-1}
\cos\bigl(\frac{2\pi m_4}{N_T}~t\bigr)~\Biggl(
\frac{128 \mu_0^2 f_{a0}^2}{[2\sin(\tfrac{1}{2}\tfrac{2\pi m_4}{N_T})]^2+m_{a0}^2}\\
&+~~4\mu_0^2~
\frac{1}{N_L^3N_T}\sum_{n_{1,2,3}=-N_L/2}^{N_L/2-1}~~\sum_{n_4=-N_T/2}^{N_T/2-1}
\biggl\{N_{f}~ \frac{1}{(k+p)^2+M_{1S}^2}~ \frac{1}{k^2+M_{2S}^2}\nonumber\\
&-\frac{1}{N_{f}}~\frac{1}{(k+p)^2+M_{12}^2}~\frac{k^2+M_{SS}^2}{1+\frac{k^2+M_{SS}^2}{N_{f} ~m_0^2/3}}~\biggl[\frac{1}{(k^2+M_{11}^2)^2}+\frac{1}{(k^2+M_{22}^2)^2}+\frac{2}{(k^2+M_{11}^2)(k^2+M_{22}^2)}\biggr]\biggl\}\Biggr)\nonumber 
\end{align}
with 
\begin{align}
k^2&=\sum_{i=1}^3 [2\sin(\tfrac{1}{2}~ \tfrac{2\pi n_i}{N_L})]^2+[2\sin(\tfrac{1}{2}~ \tfrac{2\pi n_4}{N_T})]^2\\
(k+p)^2&=\sum_{i=1}^3 [2\sin(\tfrac{1}{2}~ \tfrac{2\pi n_i}{N_L})]^2+[2\sin(\tfrac{1}{2} \{\tfrac{2\pi n_4}{N_T}+\tfrac{2\pi m_4}{N_T}\})]^2~.\nonumber
\end{align}
The notation is given in section 4. This correlator is compared 
with the lattice correlators $C^{lat}(t)$ (\ref{c_lat}) in section 5.


\newpage

\begin{table}[h]
\begin{center}
\begin{tabular}{c c c }
\hline
\hline
$m_{sea}$ & $m_{val}$ & configs.   \\ 
\hline
\hline
$0.02$ & $0.01-0.05$ & 94  \\
\hline
$0.03$ & $0.02-0.04$ & 94  \\
\hline
$0.04$ & $0.04$      & 94  \\
\hline
\hline
\end{tabular}
\end{center}

\vspace{-0.1cm}

\caption{ \small The summary of the scalar and pseudoscalar correlators analyzed in this work. All correlators are calculated at $V=16^3\times 32$,  $a^{-1}\!\approx\! 1.7$ GeV, degenerate valence quarks, two degenerate sea quarks, point source and point sink.   }\label{tab.data}
\end{table}


\begin{table}[h]
\begin{center}
\begin{tabular}{c c c c c c}
\hline
\hline
$m_{val}=m_{sea}$ & $m_{a0}$  & $f_{a0}$   & $t$  & $\chi^2$ & $dof$\\ 
\hline
$0.02$ & $0.92(9)$ & $0.044(9)$  & $4-10$ & $0.1$ & $5$ \\ 
$0.03$ & $0.99(10)$& $0.048(13)$ & $5-10$ & $0.8$ & $4$ \\
$0.04$ & $0.94(5)$ & $0.042(6)$  & $5-12$ & $0.2$ & $6$ \\
\hline
\hline
\end{tabular}
\end{center}

\vspace{-0.1cm}

\caption{ \small The $m_{a0}$ and $f_{a0}$ in lattice units 
obtained from the exponential fit to the dynamical correlators with $m_{val}=m_{sea}$. Time ranges $t=t_{min}-t_{max}$, $\chi^2$ and degrees of freedom ($dof$) in the fit are also shown. }\label{tab.dyn}
\end{table}


\begin{table}[h]
\begin{center}
\begin{tabular}{c c c c c c }
\hline
\hline
$m_{sea}$ & $m_{val}$ & $M_\pi$ & $t$    & $\chi^2$ & $dof$ \\ 
\hline
       & $0.01$ & $0.222(3)$ & $8-15$ & $0.4$     & $6$ \\
       & $0.02$ & $0.294(2)$ & $8-15$ & $0.06$    & $6$ \\
$0.02$ & $0.03$ & $0.353(2)$ & $8-15$ & $0.02$    & $6$ \\
       & $0.04$ & $0.405(2)$ & $8-15$ & $0.03$    & $6$ \\
       & $0.05$ & $0.453(2)$ & $8-15$ & $0.07$    & $6$ \\
\hline
       & $0.02$ & $0.304(2)$ & $8-15$ & $0.2$    & $6$ \\
$0.03$ & $0.03$ & $0.362(2)$ & $8-15$ & $0.05$   & $6$ \\
       & $0.04$ & $0.412(2)$ & $8-15$ & $0.02$    & $6$ \\
\hline
$0.04$ & $0.04$ & $0.408(2)$ & $8-15$ & $0.4$    & $6$ \\
\hline
\hline
\end{tabular}
\end{center}

\vspace{-0.1cm}

\caption{ \small The pion mass $M_\pi$ in the lattice units, which is obtained from the fit to the pseudoscalar correlators at various $m_{sea}$ and $m_{val}$. }\label{tab.pi}
\end{table}


\begin{table}[h]
\begin{center}
\begin{tabular}{c c c c c c c c}
\hline
\hline
$m_{val}$ &  type of fit to Eq. (\ref{c_pqchpt})  & $m_0$ [MeV] & $m_{a0}$   & $f_{a0}$   & $t$  & $\chi^2$ & $dof$\\ 
\hline
       & exponential fit           &          &            &            &      &        &    \\
$0.01$ & bubble and $a_0$-exchange & $\infty$ & $0.87(17)$ & $0.040(15)$ &$4-8$& $0.2$  & $3$\\
       & bubble and $a_0$-exchange & $800$    & $1.2(5)$   & $0.077(70)$ &     & $1.2$ & \\
       & bubble and $a_0$-exchange & $600$    & $1.8(9)$   & $0.19(30)$ &      & $3.0$ & \\
\hline
       & exponential fit           &          & $0.92(9)$  & $0.044(9)$ &      & $0.1$ & \\
$0.02$ & bubble and $a_0$-exchange & $\infty$ & $0.92(9)$  & $0.044(9)$ &$4-10$& $0.1$ & $5$\\
       & bubble and $a_0$-exchange & $800$    & $0.91(10)$ & $0.041(9)$ &      & $0.1$ & \\
       & bubble and $a_0$-exchange & $600$    & $0.92(10)$ & $0.042(10)$ &     & $0.1$ & \\  
     
\hline
       & exponential fit           &          & $0.86(4)$  & $0.040(4)$ &      & $3.0$ & \\
$0.03$ & bubble and $a_0$-exchange & $\infty$ & $0.89(5)$  & $0.041(5)$ &$4-11$& $2.0$ & $6$\\
       & bubble and $a_0$-exchange & $800$    & $0.87(5)$  & $0.039(4)$ &      & $1.8$ & \\
       & bubble and $a_0$-exchange & $600$    & $0.87(5)$  & $0.038(4)$ &      & $1.8$ & \\
\hline
       & exponential fit           &          & $0.84(4)$  & $0.035(3)$ &      &  $1.4$ & \\
$0.04$ & bubble and $a_0$-exchange & $\infty$ & $0.85(4)$  & $0.034(4)$ &$5-12$& $1.0$ & $6$\\ 
       & bubble and $a_0$-exchange & $800$    & $0.85(4)$  & $0.033(4)$ &      & $0.9$ & \\
       & bubble and $a_0$-exchange & $600$    & $0.84(4)$  & $0.033(4)$ &      & $0.9$ & \\ 
\hline 
       & exponential fit           &          & $0.87(3)$  & $0.036(3)$ &      & $1.9$ &\\
$0.05$ & bubble and $a_0$-exchange & $\infty$ & $0.88(3)$  & $0.034(3)$ &$5-12$& $1.6$ & $6$\\
       & bubble and $a_0$-exchange & $800$    & $0.87(3)$  & $0.034(3)$ &      & $1.4$ &\\
       & bubble and $a_0$-exchange & $600$    & $0.87(3)$  & $0.034(3)$ &      & $1.4$ &\\           

\hline
\hline
\end{tabular}
\end{center}

\vspace{-0.1cm}

\caption{ \small The $m_{a0}$ and $f_{a0}$ in lattice units obtained from the fit of the correlators 
with the PQChPT prediction (\ref{c_pqchpt}) at  $m_{sea}=0.02$ and various $m_{val}$. 
The fits denoted by ``bubble and $a_0$-exchange'' take into account 
both terms in Eq. (\ref{c_pqchpt}). The bubble contribution in (\ref{c_pqchpt}) depends on the value 
of $m_0$, which is taken to be $m_0=600~$MeV$,~800~$MeV$,~\infty$.   
We also present the results of the 
conventional exponential fit $e^{-m_{a0}t}+e^{-m_{a0}(N_T-t)}$, which is obtained under the assumption that the bubble contribution vanishes; 
the correlator with $m_{val}\!<\!m_{sea}$ is negative  (Fig. \ref{fig.1}) and can not be described by $e^{-m_{a0}t}+e^{-m_{a0}(N_T-t)}$.}\label{tab.4}
\end{table}

\end{document}